\documentclass{PoS}
\usepackage[authoryear,square]{natbib}
\bibpunct{(}{)}{;}{a}{}{,}

\def\ie{{\frenchspacing\it i.e.}}
\def\eg{{\frenchspacing\it e.g.}}

\def\be{\begin{equation}}
\def\ee{\end{equation}}
\def\ba{\begin{eqnarray}}
\def\ea{\end{eqnarray}}

\title{Model-independent constraints on dark energy and modified gravity with the SKA}

\ShortTitle{PCA of dark energy and modified gravity parameters}

\author{Gong-Bo Zhao $^{1,\ 2}$, David Bacon $^{2}$, Roy Maartens $^{3,\ 2}$, Mario Santos $^{3,\ 4}$, Alvise Raccanelli $^{5, \ 6, \ 7}$\\ \\

        $^{1}$ National Astronomy Observatories, Chinese Academy of Science, Beijing, 100012, People's Republic of China\\ 
        $^{2}$ Institute of Cosmology and Gravitation, University of Portsmouth, Portsmouth, PO1 3FX, United Kingdom\\ 
        $^{3}$ Department of Physics, University of the Western Cape, Cape Town 7535, South Africa\\ 
        $^{4}$ SKA SA, 4th Floor, The Park, Park Road, Pinelands, 7405, South Africa\\ 
        $^{5}$ Department of Physics \& Astronomy, Johns Hopkins University, 3400 N. Charles St., Baltimore, MD 21218, USA\\
        $^{6}$ Jet Propulsion Laboratory, California Institute of Technology, Pasadena CA 91109, USA\\ 
        $^{7}$ California Institute of Technology, Pasadena CA 91125, USA\\
     }

\abstract{Employing a nonparametric approach of the principal component analysis (PCA), we forecast the future constraint on the equation of state $w(z)$ of dark energy, and on the effective Newton constant $\mu(k,z)$, which parameterise the effect of modified gravity, using the planned SKA HI galaxy survey. Combining with the simulated data of Planck and Dark Energy Survey (DES), we find that SKA Phase 1 (SKA1) and SKA Phase 2 (SKA2) can well constrain $3$ and $5$ eigenmodes of $w(z)$ respectively. The errors of the best measured modes can be reduced to 0.04 and 0.023 for SKA1 and SKA2 respectively, making it possible to probe dark energy dynamics. On the other hand, SKA1 and SKA2 can constrain $7$ and $20$ eigenmodes of $\mu(k,z)$ respectively within 10\% sensitivity level. Furthermore, 2 and 7 modes can be constrained within sub percent level using SKA1 and SKA2 respectively. This is a significant improvement compared to the combined datasets without SKA.}

\FullConference{
Advancing Astrophysics with the Square Kilometre Array\\
June 8-13, 2014\\
Giardini Naxos, Italy}

\begin{document}

\section{Introduction}

The physical origin of the acceleration of the universe remains unknown since its discovery in 1998 using supernovae (SN) observations \citep{Riess,Perlmutter}. It might imply that there exists a repulsive `dark energy' component dominating the universe, or that we need a better understanding of the law of gravity, \ie, the general relativity (GR) might need to be modified on cosmological scales (For a recent review of modified gravity theories, see \citealt{Clifton:2011jh}). Although dark energy (DE) and modified gravity (MG) can accelerate the universe at the background level in the same way after the required tuning, the degeneracy can be broken when the cosmic structure formation is investigated.   

In this era of precision cosmology, a combination of multiple observation probes including SN, cosmic microwave background (CMB), and large scale structure (LSS) surveys is key to unveil the mystery of the cosmic acceleration \citep{Weinberg:2012es}. This is because different kinds of surveys can be highly complementary, \eg, the weak lensing (WL) and redshift surveys are able to probe $\gamma(k,z)$, quantifying the deviation of photon's trajectory from the geodesics, and $\mu(k,z)$, the time and spatial variation of the Newton's constant respectively, which are two different effects predicted by a wide range of MG models, making the combination of WL with redshift surveys robust for GR tests, as well as for the dark energy studies.   

Given that the CMB and WL surveys of Planck \citep{Planck} and Dark Energy Survey (DES) \footnote{More details of the Dark Energy Survey are available at http://www.dark-energysurvey.org/} are accumulating data, we need large redshift surveys to complement. The BOSS spectroscopic survey \citep{boss} of SDSS-III is currently the largest redshift survey worldwide, mapping the 10,000 square degree sky up to $z=0.7$ by tracing 1.5 million luminous galaxies. It will be succeeded by eBOSS \footnote{More details of the eBOSS survey are available at http://www.sdss.org/sdss-surveys/eboss/}, a multi-tracer spectroscopic survey of SDSS-IV, which will focus on a smaller patch of the sky (7500 square degree) but going deeper. According to the forecast, it will achieve 1-2\% distance measurement from the baryon acoustic oscillations (BAO) between $0.6 < z < 2.5$. The Square Kilometre Array (SKA) \footnote{More details of the SKA survey are available at https://www.skatelescope.org/} HI galaxy redshift survey can provide us with accurate redshifts (using the 21cm line) of millions of sources over a wide range of redshifts, making it an ideal redshift survey for cosmological studies \citep{SKABAO,SKARSD,SKALSST,SKAEuclid,SKAsurvey,SKAcosrev}. 

Traditionally, observational constraints on DE or MG using either current or future data are usually performed in a parameterised fashion, \ie, the equation of state of DE, $w(z)$, or the $\mu(k,z)$ and $\gamma(k,z)$ functions quantifying the effect of MG \citep{Zhao:2008bn} \footnote{There are other ways to parameterise the effect of MG, \eg, see \cite{Baker:2012zs}.}, are parameterised using assumed function forms, and then the observational constraints on these parameters are worked out. Simple as it is, this approach has its drawbacks,\begin{itemize}
\item It may cause {\it theoretical bias}: the result largely depends on the functional form used for the parametrisation, which is {\it a priori}. The functional forms are usually chosen for the purpose of simplicity, or for the assumed theoretical consistency, or for both;
\item The number of parameters are usually minimised, \eg, the CPL parametrisation \citep{CP,L} of $w(z)$ has $2$ parameters, while the BZ parametrisation \citep{BZ} for MG has $5$ parameters. This can yield a reasonably good constraint on the reconstructed $w(z)$, or the MG functions even when data is weak, but it might under fit the data when data is excellent.  

\end{itemize}

However, in nonparametric methods, including the principal component analysis (PCA), it is the assumption, rather than the number of parameters, that is minimised, hence it can largely avoid the theoretical bias. 

In this chapter, we use the PCA method to perform the forecast of $w(z)$ and $\mu(k,z)$ using a SKA HI redshift galaxy survey.

\section{Methodology}

In this section, we employ a standard Fisher matrix technology \citep{Fisher} to perform the future forecast.   

\subsection{The Fisher matrix formulism}


For a redshift survey, the Fisher matrix formalism reads \citep{FisherPk} \footnote{Note that this is the Fisher matrix for a given redshift bin. The final Fisher matrix is the sum over the Fisher matrices of individual redshift bins.},

\ba \label{eq:Fisher}F_{ij}&=&\frac{1}{8{\pi}^2}\int_{-1}^{1} {\rm d}\mu \int_{k_{\rm min}}^{k_{\rm max}}  {\rm d}k~\frac{\partial {\rm ln} \tilde{P}(k,\mu)}{\partial p_i} \frac{\partial {\rm ln} \tilde{P}(k,\mu)}{\partial p_j} V_{\rm eff}(k,\mu) k^2 e^{-k^2\mu^2\Sigma^2} \\ 
\label{eq:Kaiser}\tilde{P}(k,\mu) &=& (b+f\mu^2)^2 P(k) \\
\label{eq:Veff} V_{\rm eff}(k,\mu)&=&\left[\frac{nP(k)(1+\beta\mu^2)^2}{nP(k)(1+\beta\mu^2)^2+1} \right]^2 V_{\rm sur}
\ea where $\tilde{P}(k,\mu), V_{\rm eff}$ denote the power spectrum in redshift space and the effective volume respectively, and $V_{\rm sur}$ is the actual volume of the redshift survey. We have used the Kaiser formula, \ie, Eq (\ref{eq:Kaiser}) to evaluate $\tilde{P}(k,\mu)$, where $P(k)$ is the linear matter spectrum calculated using {\tt CAMB} \citep{camb}, $b$ and $f$ are the linear bias and the growth function respectively. To account for the Finger of God (FoG) effect, we have chosen $\Sigma$ to be $4$ Mpc, which is consistent with simulations.    

The Fisher matrix formulae for CMB and WL surveys are elaborated in \cite{Zhao:2008bn}. 

\subsection{Specifications of future SKA HI surveys}

\begin{figure}[tbp]
\centering
\includegraphics[scale=0.3]{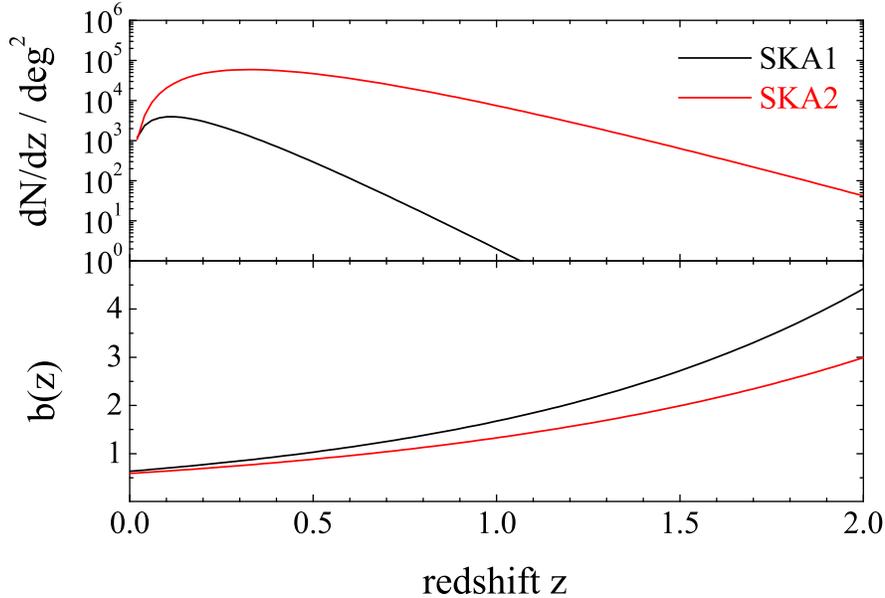}
\caption{Upper panel: The expected galaxy distribution for SKA1 and SKA2; Lower panel: the corresponding bias as a function of redshift.} \label{fig:nz}
\end{figure}

A future SKA HI redshift survey will trace the galaxies at radio wavelengths, and the redshifts will be measured precisely using the emission lines. In this work, we consider Phase 1 and Phase 2 of SKA HI surveys (dubbed SKA1 and SKA2 respectively). SKA1 will achieve an RMS flux sensitivity of $S_{\rm rms}\simeq 70 - 100 \mu$Jy with SKA1-MID or SUR, surveying over $5000$ deg$^2$ in 10,000 hours. The expected total number of galaxies in Phase 1 is roughly 5 million at redshift $z\lesssim0.5$ with a $5\sigma$ detection. In Phase 2, a 10,000 hours survey over $30,000$ deg$^2$ will detect one billion galaxies at a $10\sigma$ detection level. The expected galaxy distribution and bias for SKA1 (SKA2) is shown in Fig~\ref{fig:nz}. For more details of the survey specifications, see \cite{SKAsurvey}. Although SKA1 is not able to compete with the BOSS survey, SKA2 will surpass any planned spectroscopic surveys in the optical bands at $z\lesssim1.4$. 


\subsection{Cosmological parameters}

To be generic, we parameterise the universe using parameters \be\label{eq:para} P=\{\Omega_b h^2, \Omega_c h^2, h, \tau, n_s, A_s, w_i,\mu_{ij}, \gamma_{ij} \} \ee where $\Omega_b h^2$ and $\Omega_c h^2$ are energy density of baryons and cold dark matter respectively, $h$ is the Hubble constant, $\tau$ is the optical depth, $n_s$ and $A_s$ are the spectral index and the amplitude of the primordial power spectrum respectively. $w$ denotes the equation-of-state of dark energy. In general, we treat $w(z)$ as a unknown function and determine how many degrees of freedom of it can be constrained using the PCA method~\citep{wPCA1,wPCA2,wPCA3,MGPCA1,MGPCA2,MGPCA3}. To do this, we bin $w$ in the late-time universe, namely, $0 \leq z \leq 30$ using $M+1$ $z$-bins, and consider the value of $w$ in each bin as an independent parameter. Since the surveys we consider in this work will not be able to probe $z>3$ in detail, we use $M$ bins linear in $z$ for $0\leq z \leq 3$ and a single bin for $3 \leq z \leq 30$. 

The $\mu$ and $\gamma$'s are modified gravity parameters and they are defined as follows. 

In Newtonian gauge, the linear scalar perturbations to the flat
Friedmann-Robertson-Walker metric read,
\begin{equation}\label{FRW}
ds^2=-a^2(\eta)[(1+2\Psi(\vec{x},\eta))d\eta^2-(1-2\Phi(\vec{x},\eta))d\vec{x}^2],
\nonumber
\end{equation}
where $\eta$ is the conformal time and $a(\eta)$ the scale factor.
In Fourier space, one can write~\citep{Hu:2007pj,BZ},
\begin{eqnarray}
\label{parametrization-Poisson} k^2\Psi&=&-\mu(k,a) 4\pi G a^2\rho\Delta \\
\label{gamma}\Phi/\Psi&=&\gamma(k,a)
\end{eqnarray}
where $\Delta$ is the comoving matter density perturbation. The
function $\gamma$ describes anisotropic stresses, while $\mu$
describes a time- and scale-dependent rescaling of Newton's constant
$G$, as well as the effects of DE clustering or massive neutrinos. In $\Lambda$CDM, the
anisotropic stress due to radiation is negligible during matter
domination, thus $\mu=\gamma=1$. 

Similar to $w(z)$, we treat $\mu(k,a)$ and $\gamma(k,a)$ as unknown functions and forecast how well we can constrain the eigenmodes of them using PCA. Since they are 2-variable functions in both $k$ and $a$, we have to {\it pixelise} them in the $(k,z)$ plane. We pixelise the late-time and large-scale universe ($0 \leq z \leq 30,10^{-5} \leq k \leq 0.2~{\rm
h}\,{\rm Mpc}^{-1}$) into $M+1$ $z$-bins and $N$ $k$-bins, with each of the $(M+1)\times N$ pixels having independent values of
$\mu_{ij}$ and $\gamma_{ij}$. We consider $w(z)$ as another unknown function, allowing each of the $M+1$ $z$-bins to have an independent value of $w_i$. We use $M$ bins linear in $z$ for $0 \leq z \leq 3$ and a single bin for $3 \leq z \leq 30$. We choose $M=N=20$ and have checked that this pixelisation is fine enough to ensure the convergence of the results. We use logarithmic $k$-bins on superhorizon scales and linear $k$-bins on subhorizon scales, to optimize computational efficiency. As in~\cite{Zhao:2008bn}, we only consider information from scales well-described by linear perturbation theory, which is only a fraction of the $(k,z)$-volume probed by future surveys. Since the evolution equations~\citep{Zhao:2008bn} contain time-derivatives of $\mu(k,z)$, $\gamma(k,z)$ and $w(z)$, we follow~\cite{wPCA2} and \cite{MGPCA1} and use hyperbolic tangent functions to represent steps in these functions in the $z$-direction, while steps in the $k$-direction are left as step functions. The total number of free parameters in our forecast is therefore $(M+1)(2N+1)+17=878$.

\subsection{The principal component analysis (PCA) method}

The PCA method is a traditional method in data analysis. It helps to identify the {\it principal components} (PCs) of data by maximising the data covariance matrix. In cosmology, PCA has been used in determining the well-constrained combinations \footnote{The PCA method discussed here identifies the best measured {\it linear} combinations of cosmological parameters, although extensions exist, \eg, the kernel PCA method \citep{kPCA} can optimise the constraint on {\it nonlinear} combinations of parameters.} of cosmological parameters, \eg, the binned equation of state $w(z)$ of dark energy \citep{wPCA1,wPCA2,wPCA3} and the pixelised 2-variable functions of $\mu(k,z)$ and $\gamma(k,z)$ \citep{MGPCA1, MGPCA2, MGPCA3}, which quantify the deviation from general relativity on cosmological scales.    

Generically, the PCA method can be formulated as follows. Let $F$ be an $N \times N$ Fisher information matrix for a parameter set $P=\{p_1,p_2,...,p_N \}$. We can find the {\it eigenmodes} of $F$ by matrix diagonalisation, namely, 
\be F = W^T \ \Lambda \  W, \ee where $\Lambda={\rm diag}(\lambda_1,\lambda_2,...,\lambda_N)$, and $W$ is the transformation matrix relating $P$ to $Q$, which is a set of new parameters $Q=\{q_1,q_2,...,q_N \}$. $P$ and $Q$ are related via, \be Q=WP \ee The matrices $\Lambda$ and $W$ store the eigen-values and eigen-vectors of $F$: $W$ tells how to map the old correlated parameters, the $p$'s, to the new orthogonal ones, the $q$'s, and $\Lambda$ quantifies the uncertainty on the $q$'s. The best measured parameter is the $q$ with the minimal error (the one corresponding to the maximum entry in matrix $\Lambda$). 

For dark energy, the $p$'s are the binned $w(z)$ in redshift $z$. $W$ helps to locate the `sweet-spots' (the redshifts where the error on $w(z)$ get minimised), and $\Lambda$ quantifies how `sweet' they are (the size of errors when measuring these modes). For modified gravity, the $p$'s are the pixelised functions of $\mu(k,z)$ and $\gamma(k,z)$ in the $(k,z)$ plane, and the eigen-vectors in this case are 2D surfaces.

\section{Results}

For a given set of parameter values, we use MGCAMB~\citep{Zhao:2008bn,mgcamb} to compute the observables. We generate numerical derivatives of
observables with respect to parameters, and use the specifications
for the experiments to compute the Fisher information matrix, which
defines the sensitivity of the experiments to these parameters
(see~\citealt{Zhao:2008bn} for computational details). Our fiducial values
are in all cases $\Lambda$CDM: $\gamma_{ij} = \mu_{ij} = -w_i =
1$ for all $i$ and $j$, and the fiducial values of the other parameters are
those of Planck.

Besides the SKA HI surveys, we consider the two-point correlations (both auto- and cross-)
between weak lensing shear (WL), and cosmic
microwave background (CMB) temperature anisotropy, plus the CMB
E-mode polarization and its correlation with the CMB temperature.
Detailed descriptions of our assumptions for each measurement are
found in~\cite{Zhao:2008bn}.  WL is sourced by the sum of
the potentials $(\Psi+\Phi)$. CMB data probe the Integrated
Sachs-Wolfe effect (ISW) which depends on $\partial(\Phi+\Psi)/\partial\eta$.
Thus, measuring WL over multiple redshift bins, along with
CMB data, yields information about the relation between $\Psi$ and
$\Phi$ and their response to matter density fluctuations.
For our forecasts, we assume the following probes: Planck~\citep{Planck} for CMB, and DES for WL. 

In what follows in the section, we shall present the results of our forecast for dark energy and modified gravity respectively. 

\subsection{Dark Energy constraints}

In this section, we focus on DE constraints in the framework of general relativity. Therefore we fix the $\mu$ and $\gamma$ pixels to be unity, but vary the remaining parameters in Eq (\ref{eq:para}) simultaneously. After marginalising over other parameters, we perform a PCA on the $w$ bins.      

\begin{figure}[thp]
\centering
{\includegraphics[scale=0.4, ]{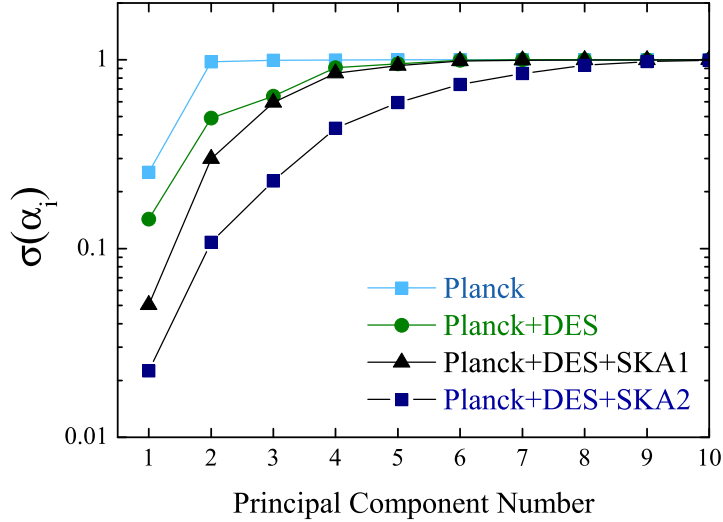}}
\caption{The forecasted 68\% CL measurement error on $\alpha_i$, the coefficient of the $i$th principal components of $w(z)+1$, namely, $w(z)+1=\sum_i \alpha_i e_i(z)$, using different data combinations illustrated in the legend. A weak prior of $\sigma(w(z))<1$ was assumed.}
\label{fig:w-eval}
\end{figure}

\begin{figure}[thp]
\centering
{\includegraphics[scale=0.4, ]{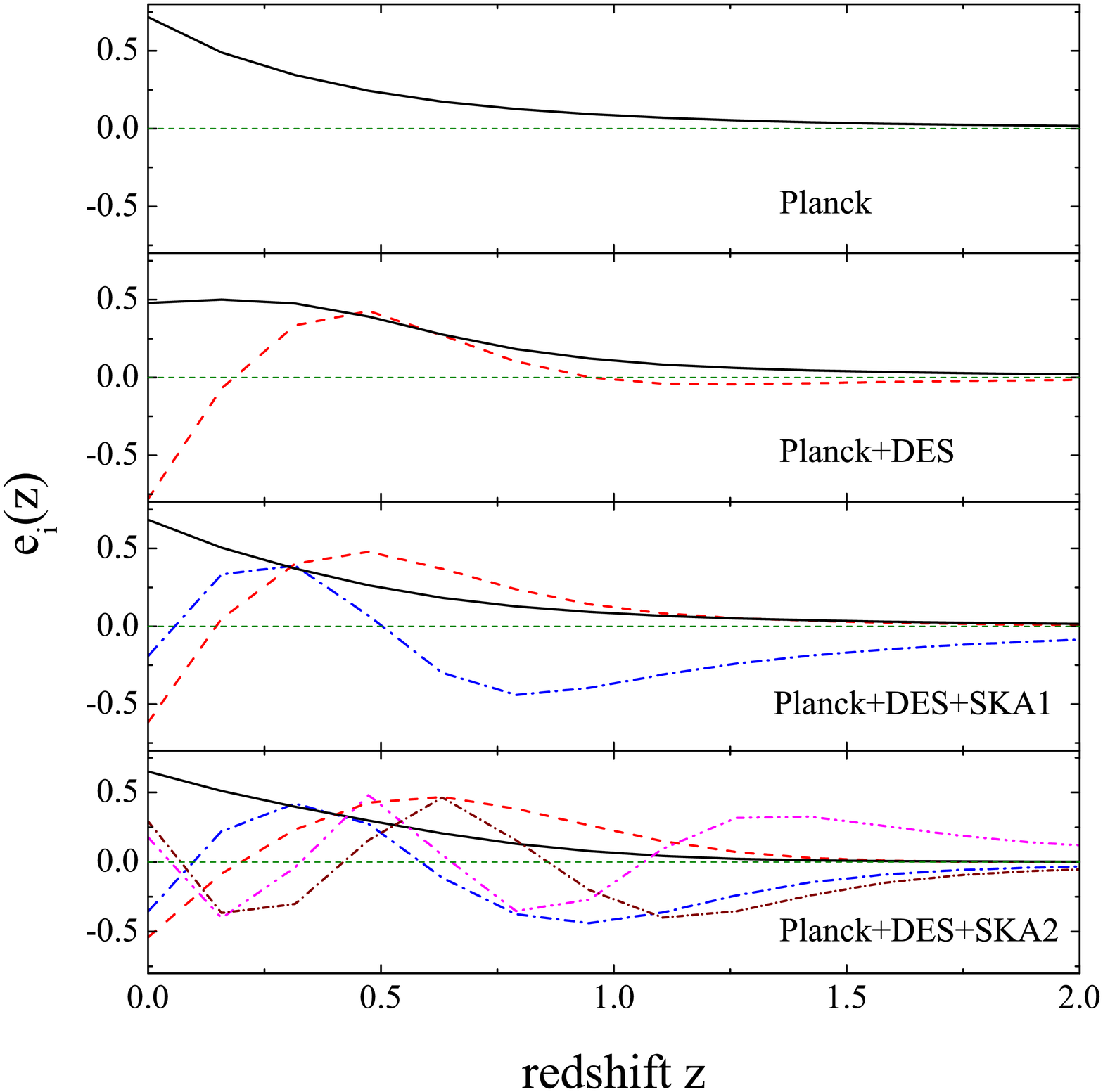}}
\caption{The best determined eigenvectors (with errors less than $0.5$) of $w(z)$ for different data combinations shown in the legends. The modes are
shown, in the order from better constrained to worse, as black solid, red dashed, blue dash-dot, purple dash-dot-dot and brown short dash-dot curves.The short dashed green horizon line shows $e_i(z)=0$. }
\label{fig:w-evec}
\end{figure}

The result is shown in Figs \ref{fig:w-eval} and \ref{fig:w-evec}. In Fig \ref{fig:w-eval}, we show the 68\% CL forecasted error on the part of the principal components (PCs) for four different data combinations. As shown, within the level of $\sigma(\alpha_i)<0.5$, Planck alone can only constrain 1 mode (the distance to the last scattering surface); Planck + DES can constrain 2 modes, while adding in SKA1 or SKA2 can constrain 3 and 5 eigenmodes to this level. In particular, the best measured modes using SKA1 and SKA2 (combined with Planck and DES) can be determined at the level of $\sigma(\alpha_1)=0.04$ and $\sigma(\alpha_1)=0.023$ respectively. This is a significant improvement given that $\sigma(\alpha_1)=0.25$ (Planck alone) and $\sigma(\alpha_1)=0.13$ (Planck + DES). 

The eigenvectors for the best constrained modes are shown in Fig \ref{fig:w-evec}. Roughly speaking, the $n$th best measured mode has $n-1$ nodes, corresponding to the $(n-1)$th time derivative of $w$. Having SKA helps determining the higher derivatives of $w(z)$, which is key to probe dark energy dynamics.

\subsection{Modified Gravity constraint}

\begin{figure}[tbp]
\centering
\includegraphics[scale=0.4]{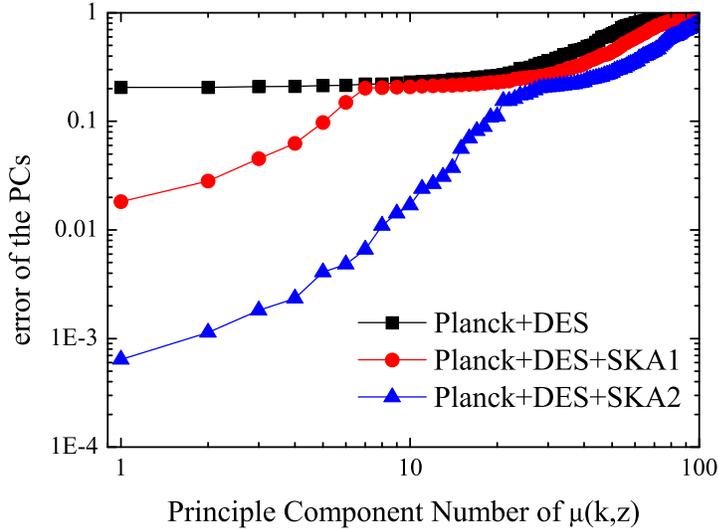}
\caption{The forecasted 68\% CL error on the coefficients of the principal components of $\mu(k,z)$ for different data combinations shown in the legend.} \label{fig:MG-eval}
\end{figure}

\begin{figure}[tbp]
\centering
\includegraphics[scale=0.5]{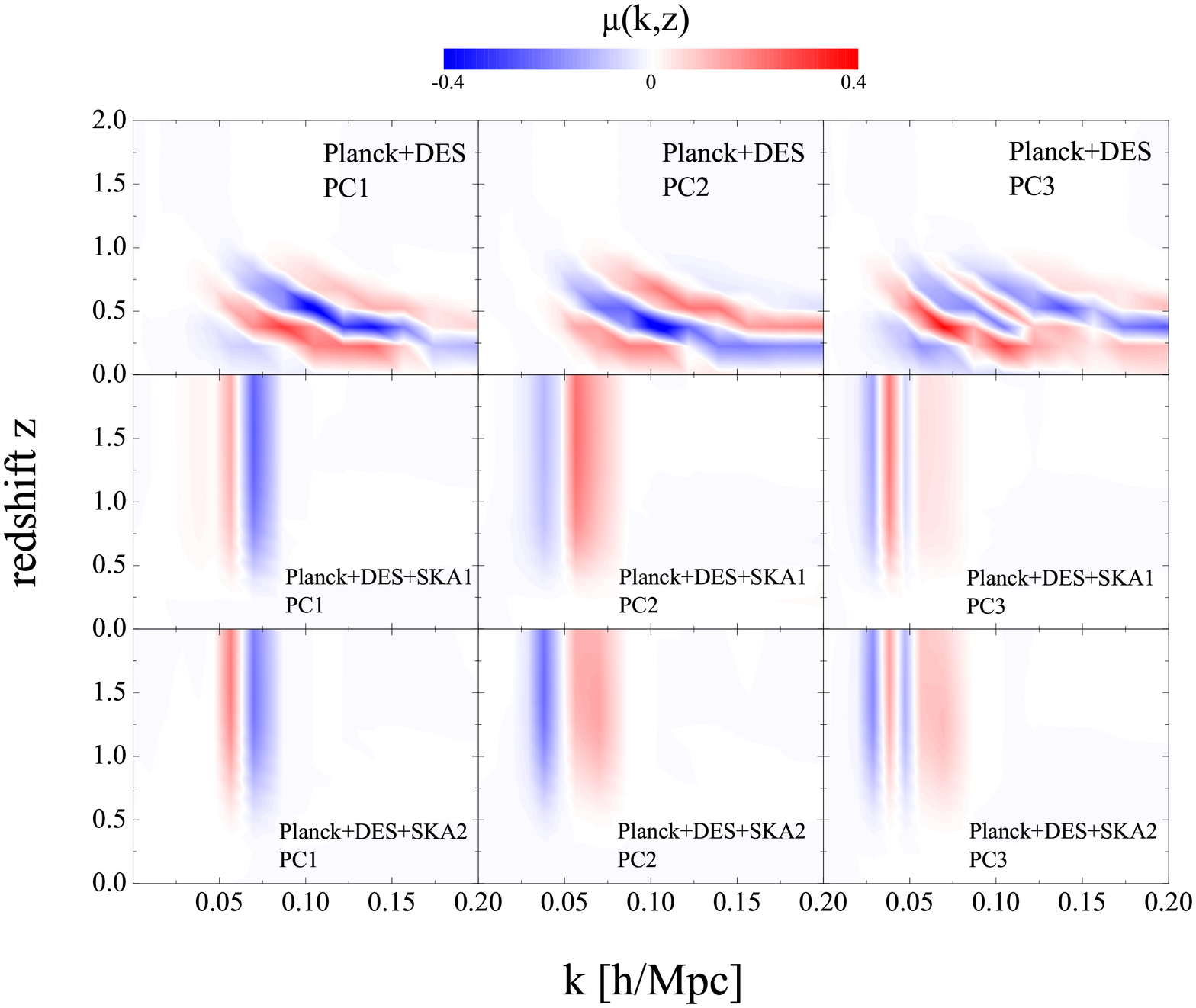}
\caption{Eigensurfaces for the first three best constrained modes of $\mu$ after marginalisation over all other cosmological parameters. Top row: Planck + DES; Middle: Planck + DES + SKA1; Bottom: Planck + DES + SKA2.} \label{fig:MG-evec}
\end{figure}

Here we consider the most general case, in which we drop the assumption of general relativity. Therefore we vary all the parameters in Eq (\ref{eq:para}) simultaneously and focus on the constraint on the $\mu$ and $\gamma$. 

Let us study the expected errors on $\mu(k,z)$ \footnote{We don't show the $\gamma$ constraint here since redshift surveys don't constrain $\gamma$ directly. }. The error on
any $\mu_{ij}$ is large, and the pixels have highly correlated
errors. We take only the $\mu_{ij}$ block of the
covariance matrix, thus marginalizing over all other parameters,
including the {$w_i$} and {$\gamma_{ij}$}. We invert this block to
obtain the Fisher matrix for our $\mu$ values, $F_{(\mu)}$, and
diagonalize $F_{(\mu)}$ by writing $F_{(\mu)}=W^{T}\Lambda{W}$. We
expect, from existing data, that variations in $\mu$ larger than
$\mathcal{O}(1)$ are unlikely. We enforce this by applying a prior
$\lambda_m>1$ to the matrix $F_{(\mu)}$. This procedure does not affect the
well-measured modes, but gives a reference point with respect to
which we define poorly constrained modes. Since we compute the full
covariance matrix, then marginalize over all but the parameter(s) of
interest, our procedure yields the results that we would get for
$\mu$ if we simultaneously measured $w$, $\gamma$, and $\mu$. This
analysis can be repeated for $\gamma$ or $w$. 

Measurements of WL and CMB probe combinations of $\Phi$ and $\Psi$, so the effects
of $\gamma$, which affects only $\Phi$, are mixed with those of
$\mu$, which affects both potentials. This yields degeneracy between
$\mu$ and $\gamma$. But this degeneracy can be broken when SKA is combined since it only measures $\Psi$ through the RSD. 

From Fig.~\ref{fig:MG-eval}, we see that Planck + DES could not constrain any modes within 10\% level, but 
adding in SKA1 can easily help to constrain 7 modes to this level. SKA2 can further increase this number to 20. In particular 2 and 7 modes can be constrained within sub percent level using SKA1 and SKA2 respectively.

Fig.~\ref{fig:MG-evec} shows three best constrained eigenmodes for $\mu$ for different data combinations. A first observation is that the modes with more nodes (a node appears when eigensurfaces crosses zero) are less constrained. This is intuitive: noisy modes are worse constrained than the smooth modes.  The best modes are mainly functions of $k$ and not $z$. This is partly because the total observable volume in the radial ($z$) direction is limited by the dimming of distant objects and, ultimately, the fact that structures only exist at relatively low $z$. Also, it is related to us considering only linear perturbations in our analysis, since at small $z$ the observable volume is too small to fit the small $k$-modes that are still in the linear regime. Hence, there is more volume available for studying the spatial distribution of structure than the radial distribution. 

For Planck+DES, we see a clear degeneracy in the $k$ and $z$ dependences of the modes. This is because 
changing $\mu$ at some point $(k,z)$ should have the same impact on the observables as a change at a larger scale but later time. Interestingly, this $k$ and $z$ dependence goes away when SKA is combined. This is simply because SKA constrains $\mu$ very well via the RSD effect, which means that data can well distinguish the effect between the variation of $\mu$ in $k$ and in $z$.

\section{Conclusion and Discussions}

In this work we apply the PCA method to investigate the constraint on dark energy and modified gravity using the future SKA HI redshift surveys, combined with CMB (Planck) and WL (DES) surveys. The PCA method is ideal to investigate dark energy and modified gravity in a nonparametric way, which efficiently minimises the theoretical bias stemming from choosing {\it ad hoc} functional forms for unknown functions.   

We study dark energy and modified gravity separately. For dark energy equation-of-state, we find that SKA Phase 1 (2) can well constrain $3$ and $5$ eigenmodes of $w(z)$ respectively. The errors on the best measured modes can be reduced to 0.04 and 0.023 for SKA1 and SKA2 respectively, making it possible to probe dark energy dynamics \citep{wrecon}. On the other hand, for modified gravity constraints, SKA1 (2) can constrain $7~(20)$ eigenmodes of $\mu(k,z)$ respectively within 10\% sensitivity level. In particular 2 and 7 modes can be constrained within sub percent level using SKA1 and SKA2 respectively. 

Imaging and redshift surveys are highly complementary when constraining cosmological parameters, especially for modified gravity models \citep{IRx1,IRx2,IRx3}. The method developed in this work can be directly applied to future surveys of LSST \citep{LSST} and Euclid \citep{Euclid}. For synergy between SKA and LSST and Euclid, see \cite{SKALSST} and \cite{SKAEuclid}.

\noindent {\bf Acknowledgement:} \\
GBZ is supported by Strategic Priority Research Program ``The Emergence of Cosmological Structures'' of the Chinese
Academy of Sciences, Grant No. XDB09000000, by the 1000 Young Talents program in China, and by the 973 Program grant No. 2013CB837900, NSFC grant No. 11261140641, and CAS grant No. KJZD-EW-T01. All numeric calculations were performed on the SCIAMA2 supercomputer at University of Portsmouth. RM and MS are supported by the South African SKA Project and the National Research Foundation. DB and RM are supported by the UK Science \& Technology Facilities Council (grant No. ST/K0090X/1). AR is supported by the Templeton Foundation. Part of the research described in this paper was carried out at the Jet Propulsion Laboratory, California Institute of Technology, under a contract with the National Aeronautics and Space Administration.

\bibliographystyle{apj}
\bibliography{PCA}

\end{document}